\documentclass[11pt]{article}

\usepackage[psamsfonts]{amssymb}
\usepackage{amsbsy}
\usepackage{times}
\usepackage{graphicx}

\makeatletter
\def\maketitle{\par 
\begingroup
   \def\thefootnote{\fnsymbol{footnote}}
   \def\@makefnmark{\hbox to 0pt{$^{\@thefnmark}$\hss}} 
   \long\def\@makefntext##1{\parindent 1em\noindent \hbox to1.8em{\hss $\m@th ^{\@thefnmark}$}##1}
   \@maketitle \@thanks
\endgroup
\setcounter{footnote}{0}
\let\maketitle\relax \let\@maketitle\relax
\gdef\@thanks{}\gdef\@author{}\gdef\@title{}\let\thanks\relax}
\makeatother

\makeatletter
\def\@maketitle{\vbox{\hsize\textwidth
\linewidth\hsize \vskip 0.1in \toptitlebar \centering
{\LARGE\bf \@title\par}  \bottomtitlebar 
   \def\And{\end{tabular}\hfil\linebreak[0]\hfil
            \begin{tabular}[t]{c}\bf\rule{\z@}{24pt}\ignorespaces}%
   \def\AND{\end{tabular}\hfil\linebreak[4]\hfil
            \begin{tabular}[t]{c}\bf\rule{\z@}{24pt}\ignorespaces}%
   \def\LINEBREAK{\end{tabular}\linebreak[4]\begin{tabular}[t]{c}\bf\rule{\z@}{16pt}\ignorespaces}%
    \begin{tabular}[t]{c}\bf\rule{\z@}{24pt}\@author\end{tabular}%
\vskip 0.3in minus 0.1in}}
\makeatother

\renewenvironment{abstract}{\vskip.075in\centerline{\large\bf Abstract}\vspace{0.5ex}\begin{quote}}{\par\end{quote}\vskip 1ex}

\makeatletter
\def\section{\@startsection {section}{1}{\z@}{-2.0ex plus -0.5ex minus -.2ex}{1.5ex plus 0.3ex minus0.2ex}{\large\bf\raggedright}}
\def\subsection{\@startsection{subsection}{2}{\z@}{-1.8ex plus-0.5ex minus -.2ex}{0.8ex plus .2ex}{\normalsize\bf\raggedright}}
\def\subsubsection{\@startsection{subsubsection}{3}{\z@}{-1.5ex plus -0.5ex minus -.2ex}{0.5ex plus .2ex}{\normalsize\bf\raggedright}}
\def\paragraph{\@startsection{paragraph}{4}{\z@}{1.5ex plus 0.5ex minus .2ex}{-1em}{\normalsize\bf}}
\def\subparagraph{\@startsection{subparagraph}{5}{\z@}{1.5ex plus  0.5ex minus .2ex}{-1em}{\normalsize\bf}}

\makeatother

\skip\footins 9pt plus 4pt minus 2pt
\def\footnoterule{\kern-3pt \hrule width 12pc \kern 2.6pt }
\leftmargini\leftmargin \leftmarginii 2em
\makeatletter
\def\@listi{\leftmargin\leftmargini}
\def\@listii{\leftmargin\leftmarginii
   \labelwidth\leftmarginii\advance\labelwidth-\labelsep
   \topsep 2pt plus 1pt minus 0.5pt
   \parsep 1pt plus 0.5pt minus 0.5pt
   \itemsep \parsep}
\def\@listiii{\leftmargin\leftmarginiii
    \labelwidth\leftmarginiii\advance\labelwidth-\labelsep
    \topsep 1pt plus 0.5pt minus 0.5pt 
    \parsep \z@ \partopsep 0.5pt plus 0pt minus 0.5pt
    \itemsep \topsep}
\def\@listiv{\leftmargin\leftmarginiv
     \labelwidth\leftmarginiv\advance\labelwidth-\labelsep}
\def\@listv{\leftmargin\leftmarginv
     \labelwidth\leftmarginv\advance\labelwidth-\labelsep}
\def\@listvi{\leftmargin\leftmarginvi
     \labelwidth\leftmarginvi\advance\labelwidth-\labelsep}
\makeatother

\belowdisplayskip \abovedisplayskip
\makeatletter
\def\normalsize{\@setsize\normalsize{11pt}\xpt\@xpt}
\def\small{\@setsize\small{10pt}\ixpt\@ixpt}
\def\footnotesize{\@setsize\footnotesize{10pt}\ixpt\@ixpt}
\def\scriptsize{\@setsize\scriptsize{8pt}\viipt\@viipt}
\def\tiny{\@setsize\tiny{7pt}\vipt\@vipt}
\def\large{\@setsize\large{14pt}\xiipt\@xiipt}
\def\Large{\@setsize\Large{16pt}\xivpt\@xivpt}
\def\LARGE{\@setsize\LARGE{20pt}\xviipt\@xviipt}
\def\huge{\@setsize\huge{23pt}\xxpt\@xxpt}
\def\Huge{\@setsize\Huge{28pt}\xxvpt\@xxvpt}
\makeatother

\def\toptitlebar{\hrule height4pt\vskip .25in\vskip-\parskip}
\def\bottomtitlebar{\vskip .29in\vskip-\parskip\hrule height1pt\vskip .09in}

\newcommand{\Eq}[1]{(\ref{#1})}

\title{A Multivariate Phase Distribution\\and its Estimation}

\author{Charles F.~Cadieu${}^*$ \& Kilian Koepsell\thanks{Both authors contributed
    equally to this work.}\\
Redwood Center for Theoretical Neuroscience \\
Helen Wills Neuroscience Institute \\
University of California, Berkeley \\
Berkeley, CA 94720 \\
\texttt{\{cadieu, kilian\}@berkeley.edu}
}

\begin{document}

\maketitle

\begin{abstract}
  Circular variables such as phase or orientation have received considerable
  attention throughout the scientific and engineering communities and have
  recently been quite prominent in the field of neuroscience. While many
  analytic techniques have used phase as an effective representation, there
  has been little work on techniques that capture the joint statistics of
  multiple phase variables. In this paper we introduce a distribution that
  captures empirically observed pair-wise phase relationships. Importantly, we
  have developed a computationally efficient and accurate technique for
  estimating the parameters of this distribution from data. We show that the
  algorithm performs well in high-dimensions (d$=$100), and in cases with
  limited data (as few as 100 samples per dimension). We also demonstrate how
  this technique can be applied to electrocorticography (ECoG) recordings to
  investigate the coupling of brain areas during different behavioral
  states. This distribution and estimation technique can be broadly applied to
  any setting that produces multiple circular variables.
\end{abstract}

\section{Introduction}
Circular variables such as phase or orientation have been used effectively for
representing complex physical phenomenon and in the analysis and processing of
signals. Countless physical systems are effectively represented using phase
variables. Coupled oscillator systems are prevalent in classical physics as a
canonical model of systems ranging from coupled pendula to coupled Josephson
junctions. Oscillator models have also been effective at describing coupled
behavior in nature: chemical reaction diffusion systems, heart-lung and
circadian rhythms, and even the coupling of firefly luminescence can all be
described with phase variables (see
e.g.~\cite{Winfree2001,Kuramoto1984,Mirollo1990,Strogatz2003}). In
engineering, phase has played a key role in signal representation. From
classical Fourier analysis to modern techniques in image representation (for
example~\cite{Daugman2001,Portilla2000,Felsberg2001}), phase provides a useful
representation.

Within neuroscience, oscillatory dynamics and phase variables have had an
especially interesting history. Oscillatory dynamics played a central role in
many early theories of large-scale brain dynamics~\cite{Freeman1975}, and
oscillatory dynamics have recently received widespread
interest~\cite{Fries2005,Sejnowski2006,nips2007}. Network oscillations are
hypothesized to be functionally involved in a wide range of tasks, such as
representing sensory information, regulating the flow of information, {\em
  binding} of distributed information, and establishing and recalling
memories. Clearly, phase is of central importance to the field of
Neuroscience.

\begin{figure}[ht]
\begin{center}
\includegraphics[width=.9\linewidth]{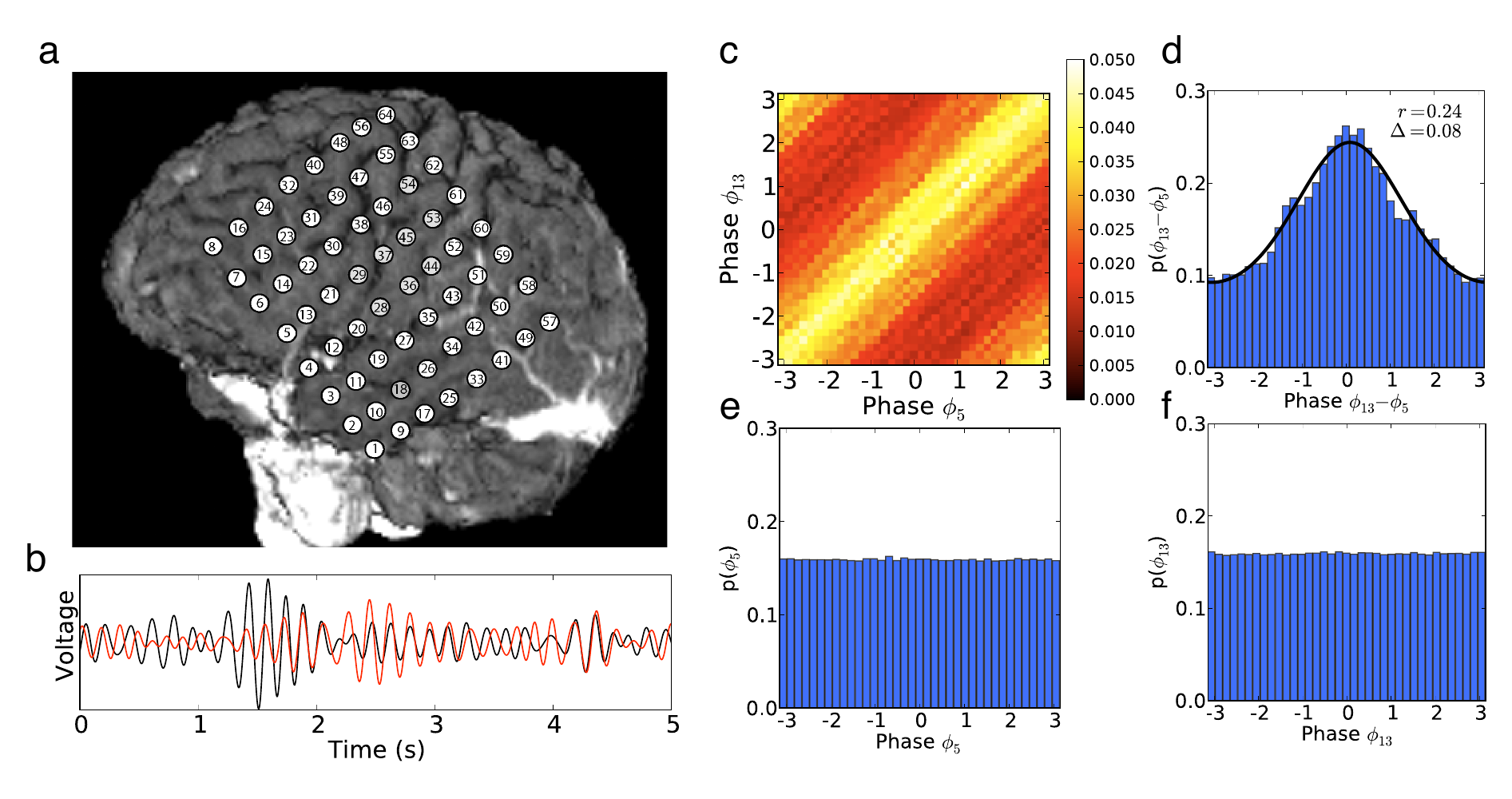} 
\end{center}
\caption{Phase dependencies in theta oscillations of human ECoG
  recordings. \textbf{a}~An 8x8 ECoG grid on the surface of cortex;
  \textbf{b}~recordings from 2 sites (5 and 13) filtered in the theta band
  ($6\pm1.2$ Hz); \textbf{c}~empirical joint phase distribution of sites 5 and
  13: note the strong dependency between the phases. \textbf{d}~The empirical
  distribution is highly concentrated in the difference of the phases,
  corresponding to phase correlations parameterized by $r
  e^{j\Delta_{ik}}\!=\langle e^{\theta_k}e^{-\theta_i}\rangle$, while the
  marginals, \textbf{e}~and \textbf{f}, are flat.}
\label{fig:ecog}
\end{figure}

In many of these cases, the relationships between phase variables is of
central interest and importance. For example, if we examine the pair-wise
relationships of phase variables recovered from electrocorticography (ECoG)
recordings we see strong statistical dependencies (see Figure~\ref{fig:ecog}
and section~\ref{sec:observations} for a more detailed discussion). The
coupling of these oscillations may indicate common sources of input or task
based cortico-cortical communication. In order to address these scientific
questions we need analytical techniques that can capture the observed
statistical dependencies.

While the need for techniques that model multidimensional phase variables may
be clear, we know of no models or techniques that provide an adequate
probabilistic representation or an effective model estimation technique. In
the real-valued case the multivariate Gaussian distribution, and in the binary
case the Ising model, serve as widely used multivariate distributions. There
is no such equivalent for phase variables (see Discussion). Because of the
lack of an appropriate probabilistic framework, many efforts have turned to
measures of phase offset and phase correlation, which we will show are
inadequate and may lead to false conclusions. Finally, the few efforts that do
apply probabilistic models to phase variables restrict themselves to low
dimensions, for example just 1 or 2 dimensions, or do not offer the
flexibility necessary to deal with the distributions we have observed
empirically.

In this paper we provide a solution to the probabilistic modeling of
multivariate phase variables and the estimation of the model distribution from
observations. We begin with a motivating example: we examine empirically
observed phase distributions recorded from a 64-channel ECoG grid analyzed for
theta band oscillations. We then introduce a multivariate phase distribution
that is capable of capturing the empirically observed distributions. Next we
provide an algorithm for recovering the parameters of our model based on
score-matching. To demonstrate the distribution and algorithm we examine a
weakly coupled oscillator system with 3 nodes. We then investigate the
performance of the algorithm as a function of dimensionality of the phase
space (2-100 dimensions) and the number of observations. As a final example,
we demonstrate how this technique could be used to model the statistical
distribution of theta phase from observations of an ECoG grid and the recovery
of the network couplings under different behavioral states.

\section{The Multivariate Phase Distribution}
\subsection{Observations from Empirical Phase Distributions}
\label{sec:observations}
Scientists and engineers have observed dependencies among phase variables in a
variety of settings~\cite{Portilla2000,Kuramoto1984,Rosenblum2001}. Here we
examine a case of direct interest to neuroscientists: phase distributions of
theta-band oscillations recorded from human patients using
electrocorticography (ECoG). In the experimental setup (for further details
see Ref.~\cite{Canolty2006}), an 8x8 electrode grid, shown in
Figure~\ref{fig:ecog}a, is placed on the surface of the cortex of an awake
behaving human patient who is receiving treatment for epilepsy. A number of
papers have described interesting phenomena of these signals, especially when
examining bandpass filtered responses. While multiple peaks are typical in the
frequency response of ECoG recordings, here we examine the theta band, which
has been implicated in cognitive processing. A common and useful technique
uses the Hilbert transform to extract an amplitude and phase from theta-band
filtered time series. Figure~\ref{fig:ecog}b displays the time series of two
simultaneous theta-band filtered responses.

When we examine the empirical phase distributions we see clear dependencies
among theta-band phase variables. The phase variables recorded from individual
ECoG sites are uniformly distributed between $-\pi$ and $\pi$, see
Figure~\ref{fig:ecog}c. However, when we examine pairs of phase variables
there are strong dependencies. In Figure~\ref{fig:ecog}d we display the
empirical joint phase distribution of two neighboring ECoG sites (sites 5 and
13). The distribution exhibits a clear diagonal structure indicating that the
probability of one phase conditioned on another is highly peaked.

This type of pairwise dependency can be described compactly using a von Mises
distribution in the difference of the phase variables. The phase distribution
of $\theta_1 - \theta_2$ can be written as:

\begin{equation}
P ( \theta_1 - \theta_2 | \kappa_{12},\mu_{12}) = \frac{1}{2\,\pi I_0(\kappa_{12})}e^{\displaystyle \kappa_{12} \cos (\theta_1 - \theta_2 - \mu_{12}) },
\label{eq:distribution2d}
\end{equation}
where $\kappa_{12}$ indicates the degree of concentration in the distribution,
$\mu_{12}$ indicates the mean phase offset between the variables, and
$I_0(\kappa)$ is the modified Bessel function of zeroth order (solid line in
Figure~\ref{fig:ecog}d). If we examine the empirical phase difference
$(\theta_1 - \theta_2)$, shown in Figure~\ref{fig:ecog}d, we see that this
type of distribution captures the variation quite well. Importantly, we have
observed similar dependencies among many neighboring ECoG sites as well as
ECoG sites separated by centimeters on the cortex. In practice, we would like
to know: which electrodes show strong phase dependencies, the strength of
these dependencies during different behavioral tasks, and the phase offset
between phase variables. While this pairwise model exhibits the appropriate
structure for two variables, it does not specify the full joint
distribution. In the next section we present a model that is capable of
capturing the full joint distribution of all 64 electrode sites.

\subsection{A Model of Multidimensional Phase Variables}
\label{sec:distribution}
Motivated by observations of empirical data, we introduce the following
$d$-dimensional multivariate phase distribution
\begin{equation}
  p(\boldsymbol{\theta}|\mathbf{K}) =
   \frac{1}{Z(\mathbf{K})} \exp[-E(\boldsymbol{\theta};\mathbf{K})] = 
  \frac{1}{Z(\mathbf{K})} \exp\left[
    {\textstyle \frac{1}{2}}\, \mathbf{x}^\dagger \mathbf{K}  \mathbf{x}
  \right]
\label{eq:distribution}
\end{equation}
where $\mathbf{x}$ is the $d$-dimensional complex vector with components
$x_i=e^{j\theta_i}$, $\mathbf{K}$ is the $d\!\times\!d$-dimensional hermitian
positive-definite matrix with elements $K_{ik} = \kappa_{ik}e^{j \mu_{ik}}$
and $Z(\mathbf{K})$ is the normalization constant needed to assure that the
probability integrates to one. Note that it is non-trivial to compute the
normalization $Z(\mathbf{K})$.

We can expand the energy function, $E(\boldsymbol{\theta};\mathbf{K})$ in
Equation~\Eq{eq:distribution} to obtain a possibly more intuitive form:
\begin{equation}
  E(\boldsymbol{\theta};\mathbf{K}) = -\frac{1}{2} \sum_{i=1}^d \sum_{j=1}^d \kappa_{ij}\cos(\theta_i-\theta_j-\mu_{ij})
\label{eq:energy}
\end{equation}

\subsection{Model Estimation}
\label{sec:fitting}

In order to fit the parameters of the distribution to a given data
distribution, we use the score-matching method introduced by
Hyvarinen~\cite{Hyvarinen2005,Hyvarinen2007}. Score-matching allows the
fitting of probability distributions of the form $p(\mathbf{x}) =\frac{1}{Z}
\exp[-E(\mathbf{x})]$, for real-valued data ($\mathbf{x}\in\mathbb{R}^n$),
without computation of the normalization constant $Z$. The values of the
parameters are obtained by minimizing a score function that contains the
log-derivative of the model density but not its normalization constant. If the
energy, $E(\mathbf{x})$, depends linearly on the model parameters, the
solution can be computed in closed form by setting the score function to
zero~\cite{Hyvarinen2007}. We follow this approach to estimate the model
parameters for our distribution~\Eq{eq:distribution}. The score matching
estimator of $\mathbf{K}$ is given by $ \widehat{\mathbf{K}} = \arg
\min_\mathbf{K} J_{\mathrm{SM}}(\mathbf{K}) $ and the score function
$J_{\mathrm{SM}}(\mathbf{K})$ is given by
\[
J_{\mathrm{SM}}(\mathbf{K}) = \left\langle \textstyle{\frac{1}{2}}
[\nabla_{\boldsymbol{\theta}}E(\boldsymbol{\theta};\mathbf{K})]
[\nabla_{\boldsymbol{\theta}}E(\boldsymbol{\theta};\mathbf{K})]^T
- \nabla_{\boldsymbol{\theta}}^2E(\boldsymbol{\theta};\mathbf{K})
\right\rangle
\]
with the expectation value, $\langle\ldots\rangle$, taken over the data
distribution. Using the quadratic form of the energy in~\Eq{eq:distribution}
and the Jacobian $D_{ij}:=\partial x_i/\partial\theta_j$, we compute
\begin{eqnarray}
\nabla_{\boldsymbol{\theta}}E(\boldsymbol{\theta};\mathbf{K})
&=& -\textstyle{\frac{1}{2}} \mathbf{x}^\dagger\mathbf{K}\mathbf{D}
   -\textstyle{\frac{1}{2}} \mathbf{D}^\dagger\mathbf{K}\mathbf{x}\nonumber\\
   \nabla_{\boldsymbol{\theta}}^2 E(\boldsymbol{\theta};\mathbf{K})
&=& \mathbf{x}^\dagger\mathbf{K}\mathbf{x}
   -\mathrm{Tr} (\mathbf{D}^\dagger\mathbf{K}\mathbf{D}) =
   -2\,E(\boldsymbol{\theta};\mathbf{K})\nonumber\\
{}[\nabla_{\boldsymbol{\theta}}E(\boldsymbol{\theta};\mathbf{K})]
{}[\nabla_{\boldsymbol{\theta}}E(\boldsymbol{\theta};\mathbf{K})]^T
&=& \textstyle{\frac{1}{2}} \mathbf{x}^\dagger\mathbf{KK}\mathbf{x}
   +\textstyle{\frac{1}{4}} \mathbf{x}^\dagger\mathbf{KD}\mathbf{D}^T\mathbf{K}^T
\mathbf{x}^*
   +\textstyle{\frac{1}{4}} \mathbf{x}^T\mathbf{K}^*\mathbf{D}^*\mathbf{D}^\dagger
\mathbf{K}\mathbf{x}
\nonumber
\end{eqnarray}
The estimator $\widehat{\mathbf{K}}$ is computed by setting the derivative of
the score function $\partial/\partial K_{ij}\,J_{\mathrm{SM}}(\mathbf{K})$ to
zero. This produces a system of linear equations:
\begin{equation}
\frac{1}{2}\sum_{k,l=1}^{d}  \big[
  \delta_{jl}\langle x_ix_k^* \rangle 
 +\delta_{ik}\langle x_lx_j^* \rangle
 -\delta_{jk}\langle x_ix_l x_j^* x_k^* \rangle
 -\delta_{il}\langle x_ix_l x_j^* x_k^* \rangle \big]\, K_{kl} 
= 2\left\langle x_i x_j^*  \right\rangle
\label{eq:linearsystem}
\end{equation}
which we can solve using standard techniques.

\section{An Example: Three Weakly Coupled Oscillators}
To further illustrate the proposed model and to demonstrate how model
parameters are recovered from a simulated system, we will work with a model of
weakly coupled oscillators. We formally introduce a system of $d$ oscillators
with kinetic energy $T\!=\!\frac{\scriptstyle 1}{\scriptstyle
  2}\,\boldsymbol{\omega}^T\mathbf{I}\,\boldsymbol{\omega}= \frac{\scriptstyle
  1}{\scriptstyle 2}\,\boldsymbol{\ell}^T\mathbf{I}^{-1}\boldsymbol{\ell}$,
where $\boldsymbol{\omega}=\boldsymbol{\dot{\theta}}$ is the vector of angular
velocities, $\boldsymbol{\ell} = \mathbf{I}\,\boldsymbol{\omega}$ are the
angular momenta, and $\mathbf{I}$ is a diagonal matrix of moments of inertia.
If we introduce a coupling between any two oscillators by a force of the form
$F_{ij}(\theta_i,\theta_j)$, the system has the Hamiltonian
$H(\boldsymbol{\theta},\boldsymbol{\ell}) =
E(\boldsymbol{\theta})+T(\boldsymbol{\ell})$.  The Hamiltonian equations
$\nabla_{\boldsymbol{\ell}}H=\boldsymbol{\dot{\theta}}$ and
$\nabla_{\boldsymbol{\theta}}H=-\boldsymbol{\dot{\ell}}$ are given as:
\begin{equation}
\boldsymbol{\dot{\theta}}=\mathbf{I}^{-1}\,\boldsymbol{\ell}\,,\qquad
\boldsymbol{\dot{\ell}}=-\mathbf{I}^{-1}\,\mathbf{F}\,.
\end{equation}
The first equation is the definition of angular momenta and the second
equation is the equation of motion. The force, $F_i=\partial/\theta_iH$,
experienced by an oscillator $i$ can be computed from~\Eq{eq:energy} as
\begin{equation}
F_i = \sum_{j=1}^d \kappa_{ij}\sin(\theta_i -\theta_j -\mu_{ij})\,.
\end{equation}
Integrating the second order equation of motion once and setting all moments
of inertia to one, we obtain the following equations of motion that we use to
simulate the system of coupled oscillators using different values for initial
angular velocities $\boldsymbol{\omega}_0$:
\begin{equation}
  \dot{\theta}_i(t) = \omega_{0i} + \sum_{j=1}^n  \int_{t'=0}^t\, \kappa_{ij}\sin(\theta_i(t') -\theta_j(t') -\mu_{ij}) \,\mathrm{d}t'\,.
\label{eq:eom}
\end{equation}
This system is similar to the Kuramoto model~\cite{Kuramoto1984}, given by
$\dot{\theta}_i\!=\omega_{0i} +
\sum_{j=1}^d\!s_{ij}\sin(\theta_i\!-\!\theta_j)$, which does not have inertia.

\begin{figure}[ht]
\begin{center}
\includegraphics[width=5.5in]{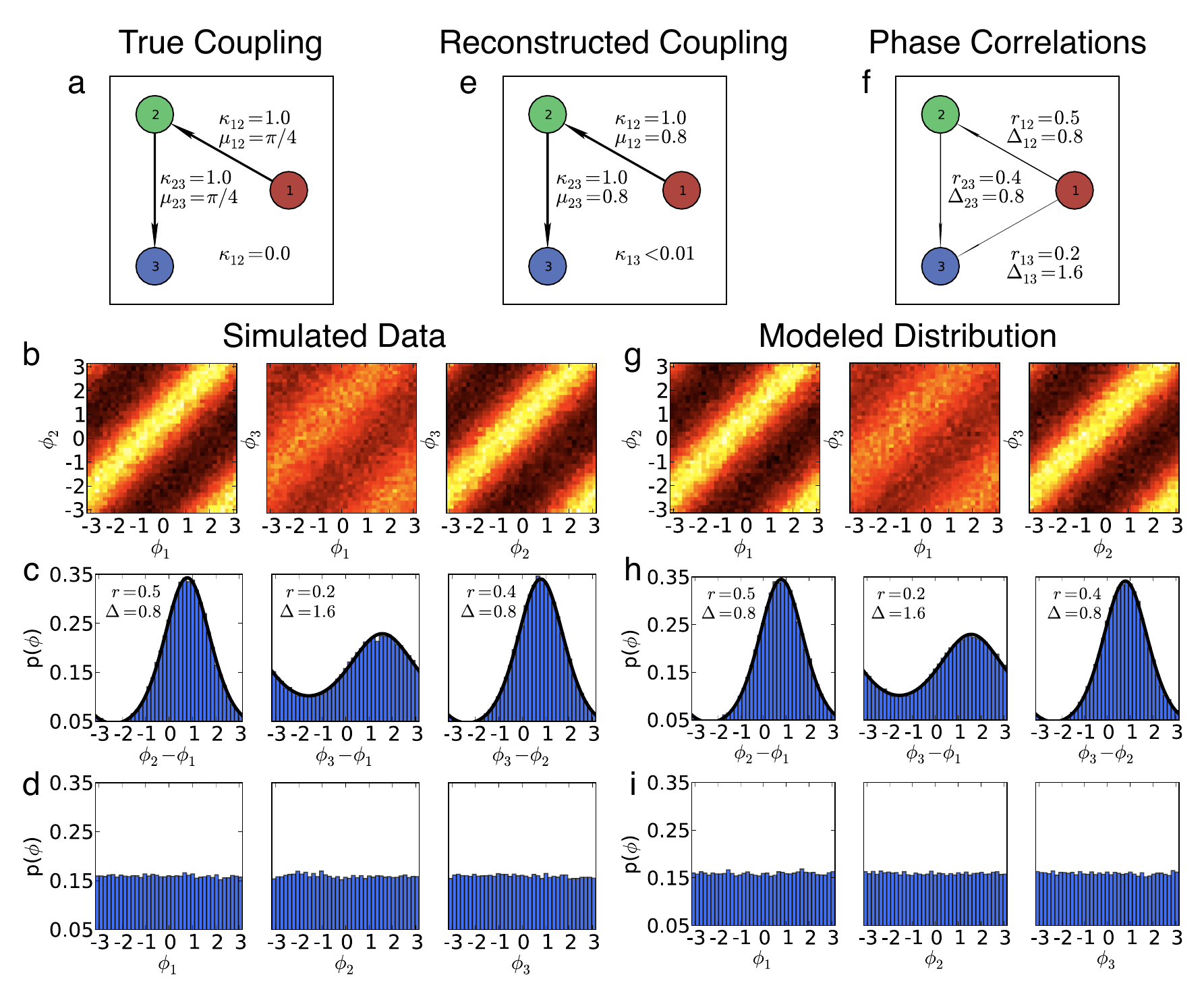}
\end{center}
\caption{Modeling a network of weakly coupled oscillators. We simulated a
  three oscillator system \textbf{a} to produce a distribution of empirical
  phase variables: \textbf{b}~pair-wise joint distributions, \textbf{c}
  pair-wise phase differences, and \textbf{d} marginal
  distributions. \textbf{e} Phase correlations calculated from the empirical
  distribution. \textbf{f}~Model parameters recovered by our technique. Using
  hMC we generated samples from the recovered distribution:
  \textbf{g}~pair-wise joint distributions, \textbf{h}~pair-wise differences,
  \textbf{i}~marginals.}
\label{fig:toymodel}
\end{figure}

For this specific example we set up a system of three weakly coupled
oscillators in which the first oscillator is coupled to the second and the
second to the third (no coupling between the first and the third). The
arrangement is shown graphically in Figure~\ref{fig:toymodel}; the direction
of the arrows indicate phase angles of the bi-directional couplings). When we
simulate this system, we observe distributions similar to those observed in
ECoG data (Figure~\ref{fig:ecog}). For example, the marginal distributions in
each of the phase variables is flat, and there are strong concentrations in
the differences of the phase variables. Interestingly, even though there is no
explicit coupling term between the first and third variables there is a strong
concentration in their difference. If we use a measure of phase correlation,
$\langle e^{\theta_k}e^{-\theta_i}\rangle = r_{ik}e^{j\Delta_{ik}}$, we see a
significant correlation between the first and third phase variables (phase
correlations are depicted in Figure~\ref{fig:toymodel}b).

The estimation technique is able to recover the underlying coupling
relationships from measurements of the simulated system\footnote{We obtained
  qualitatively similar results from simulations of the Kuramoto
  model~\cite{Kuramoto1984}.}. Simulating the system in~\Eq{eq:eom} we obtain
100,000 samples of the 3-d phase space. We then solve the linear system of
equations given by \Eq{eq:linearsystem} to estimate a $K$ matrix. The terms in
the estimated $K$ matrix are depicted in Figure~\ref{fig:toymodel}c. The
estimation technique is able to recover the parameters to within 1\% of their
true values. Importantly, the estimated parameters indicate that there is no
coupling between the first and third values (where measures of phase
correlation indicate a high concentration).

After estimating the values of the $K$ matrix we can compare the probability
distribution produced by these parameters to the empirical distribution
produced from the oscillating system. However, because the normalization
constant in equation \Eq{eq:distribution} is intractable we must use Monte
Carlo methods to estimate the true probability. While a number of techniques
may be used, we have found that the hybrid Monte Carlo (hMC) algorithm works
well for this distribution \cite{Neal1993, MCMCStuff}. We used 50 leapfrog
steps and adjusted the step size to achieve an acceptance rate close to
90\%. Taking 10,000 samples from a hMC chain we observe that the estimated
distribution closely matches the empirical distribution of the coupled
oscillator system. Again, the marginal phases are flat and there are
concentrations in the difference of the phase variables. Importantly, the
concentration of the unimodal distributions are close as are the offsets.

\section{Performance of Estimation Technique}
We now demonstrate the consistency of our estimation technique: the ability of
the technique to recover the model parameters from data. The procedure is as
follows. We begin by sampling a set of model parameters $K$. Given these model
parameters we then sample phase variables from that model using hMC. We then
estimate the model parameters given the sampled data. The real and imaginary
entries of the complex matrix $K$ are sampled from a normal distribution:
$\mathrm{Re} \{K_{ij} \}, \mathrm{Im} \{K_{ij} \} \sim N(0,1)$ and the
diagonal entries are set to zero: $K_{ij} = 0$. Note that this produces a
dense coupling matrix.

\begin{figure}[ht]
\begin{center}
\includegraphics[width=5.5in]{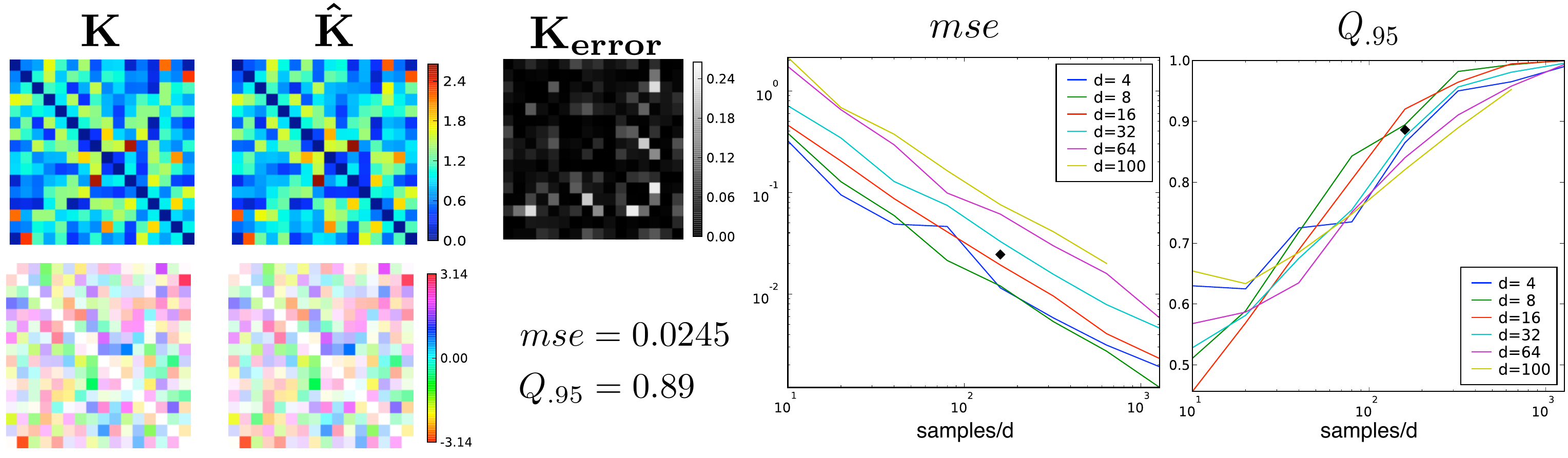} 
\end{center}
\caption{Recovery of model parameters. First column, $\mathbf{K}$: true model
  parameters for $d=16$ (first row, element-wise amplitude; second row,
  element-wise angle with alpha scaled by the amplitude). Second column,
  $\mathbf{\hat K }$: estimated parameters recovered from $2560$
  samples. Third column: element-wise $mse$ (note scaling). Fourth column:
  $mse$ metric as a function of samples per dimension for various
  dimensions. Fifth column: $Q_{.95}$ metric. The example displayed in the
  first 3 columns is indicated by the black diamond.}
\label{fig:performance}
\end{figure}

In the first column of Figure~\ref{fig:performance}, we graphically display
the element-wise amplitude and phase of a sample matrix $K$ where
$d\!=\!16$. Given this matrix we sampled $N\!=\!2560$ points using hMC. The
recovered model parameters are shown in the second column of
Figure~\ref{fig:performance}. While it is clear that these matrices are
visually similar, we quantified the error using two different metrics. First
we calculate the mean-squared-error of the matrix elements $mse =
\frac{1}{2d^2}\sum_{ij}{ ( \mathrm{Re}\{ K_{ij}\!-\!\hat K_{ij}\} )^2 +
  (\mathrm{Im}\{ K_{ij}\!-\!\hat K_{ij}\} )^2}$, where $\hat K_{ij}$ is the
estimated model parameters. In the third column of
Figure~\ref{fig:performance}a we display the element-wise error before
averaging. We also computed a metric indicating the quality of the recovered
parameters borrowed from Ref.~\cite{Timme2007}: $Q_{.95} = \frac{1}{2}\langle
u(1\!-.95 -\mathrm{Re} \{ \Delta K_{ij} \}) + u(1\!-.95 -\mathrm{Im} \{ \Delta
K_{ij} \}) \rangle_{i>j} $, where $\Delta K_{ij} = | K_{ij} - \hat
K_{ij}|/2K_{\max} $, $u$ is the Heaviside step function, and $K_{\max}$ is the
maximum absolute value of all matrix entries $K_{ij}$ and $\hat{K}_{ij}$. For
the example in Figure~\ref{fig:performance}, $mse = 0.0245$, and $Q_{.95} =
0.89$.

We computed these error metrics over a range of dimensions and samples per
dimension. The error metrics for each dimension and samples per dimension were
averaged over 10 trials and are plotted in Figure~\ref{fig:performance}b. The
algorithm achieves highly accurate model recovery for as few as 100 samples
per dimension and achieves full recovery of parameters as the number of
samples per dimension reaches 1000. This indicates that the recovery of model
parameters is quite feasible in many real world settings.

\section{A Simulation: Recovery of Couplings from Artificial 8x8 Grid Data}
A number of experimental techniques that measure large-scale cortical dynamics
may benefit from an analysis using the multivariate phase distribution and
estimation technique described in this paper. Some of the longstanding
questions within neuroscience, which may be addressed using these experimental
techniques, are ``how are neural networks coupled?'' and ``how does neural
coupling change during different behavioral conditions?''. With these
experimental questions in mind, we examine $d\!=\!64$ sampled data to
demonstrate how the multivariate phase distribution and the estimation of its
parameters can be used to recover behavioral-state-dependent brain
activity. We use a simulation rather than real ECoG recordings in order to
test our methodology while knowing the ground truth.

\begin{figure}[ht]
\begin{center}
\includegraphics[width=.9\linewidth]{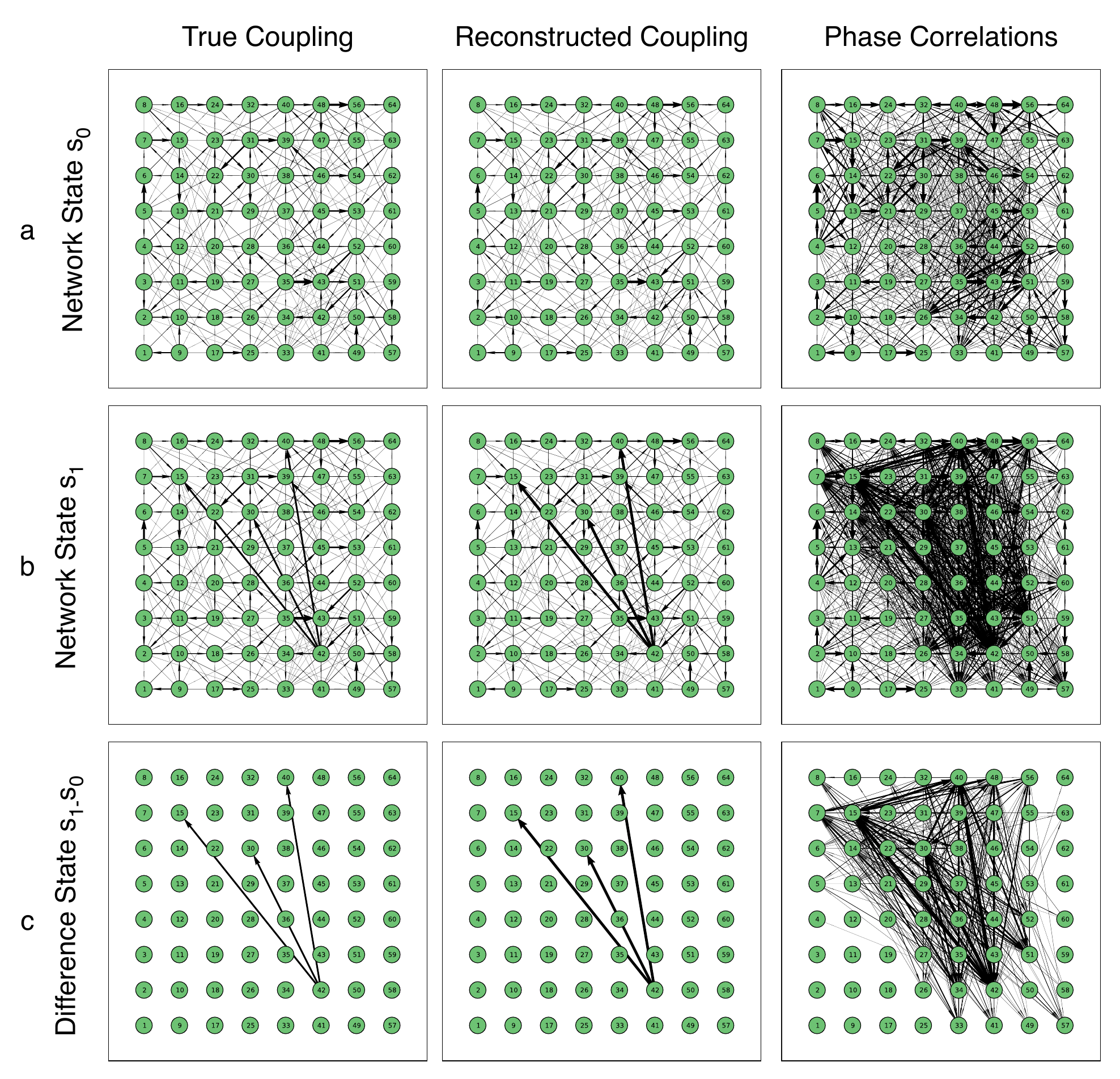} 
\end{center}
\caption{Recovery of coupling states from a simulated ECoG recording. We
  generated two network states (rows \textbf{a} and \textbf{b}). The second
  state included 3 additional couplings not present in the first state, as
  seen in the difference of states (row \textbf{c}). Taking 100,000 samples
  from each brain state, we used our algorithm to recover the model parameters
  (column 2). For comparison, the empirical phase correlations are displayed
  in column 3. Thickness of arrows indicates the magnitude of the couplings
  and directions indicate phase angles (coupling is bi-directional).}
\label{fig:gridmodel}
\end{figure}

In this simulation we used an 8x8 grid of recording sites and imposed two
different coupling matrices corresponding to two distinct brain states. The
goals are to 1) recover the couplings between recording sites during both
brain states and 2) isolate the change in the network between the two
states. We sampled a single coupling matrix, $K_{s0}$, generated from the
following distribution: the $\mu_{ij}$ terms were centered around small
offsets, and the $\kappa_{ij}$ terms were sampled from a one sided Gaussian
distribution with a variance determined by the distance between the recording
sites. The resulting coupling matrix is presented graphically in
Figure~\ref{fig:gridmodel}a. For the second simulated brain state, $K_{s1}$,
we took the original brain state, $K_{s0}$ and introduced 3 additional long
range couplings, depicted in Figure~\ref{fig:gridmodel}b. We simulated the
system using a hMC chain and took 100,000 samples from each brain state. We
then estimated the coupling matrixes for each state, $\hat K_{s0}$ and $\hat
K_{s1}$. The resulting couplings are depicted in Figures~\ref{fig:gridmodel}c
and \ref{fig:gridmodel}d. For comparison we also computed the phase
correlation measured under each brain state (depicted in the third column of
Figure~\ref{fig:gridmodel}).

This method successfully uncovers the couplings under the different brain
states and can be used to isolate the additional coupling produced by the
second brain state. The recovered couplings are depicted in the second column
of Figure~\ref{fig:gridmodel}. The parameter-estimation error metrics are:
$mse(K_{s0}, \hat K_{s0})\!=\!0.00008$, $Q_{.95}(K_{s0}, \hat
K_{s0})\!=\!1.0$, $mse(K_{s1}, \hat K_{s1})\!=\!0.0003$, and $Q_{.95}(K_{s1},
\hat K_{s1})\!=\!1.0$. By taking the difference of the $K$ matrices we can
isolate the additional coupling (shown in the third row of
Figure~\ref{fig:gridmodel}). Clearly, the difference uncovers the 3 additional
long range interactions that we introduced. In contrast, the measurement of
phase correlation produces spurious results that might lead to false
conclusions. In many instances the calculation of phase correlation produces
high values when the true concentrations are not present (see
Figure~\ref{fig:toymodel} for a simple example); taking the difference of the
correlation measurements results in a dense set of connections, in which all
but 3 are spurious.

\section{Discussion}
Our multivariate phase distribution and estimation technique can be compared
to a number of previous efforts to characterize statistical dependencies in
circular phase variables. We can break down the previous efforts by analyzing
3 major differences. First, many previous approaches only provide measurements
such as means or phase correlations and do not produce a true probabilistic
distribution over the phase variables. Second, previous examples of
probabilistic models of phase variables only characterize univariate or
bivariate distributions (see e.g.~\cite{Jammalamadaka2001,Mardia2000}). Models
similar to ours have not been extended to dimensions beyond
$d\!=\!2$~\cite{Gatto2007,Mardia2007}. Third, common multivariate phase
distributions do not capture the distributions we have observed
empirically. Most notably the von Mises-Fisher distribution only captures a
unimodal phase distribution on the hyper-sphere, which does not produce
unimodal concentrations in the differences of phase variables.

It is important to point out that our model does not model unidirectional
interactions. Because our model only captures the instantaneous distribution,
it is blind to time reversal and cannot model directed interactions. However,
extending the analysis to include multiple time slices may provide an
extension capable of modeling directed interactions (see~\cite{Rosenblum2001}
for a $d\!=\!2$ example). Closely related to modeling directed interactions is
the study of causality. Like all statistical models, the distribution and
model we have introduced here does not directly address the issue of
causality. However our distribution and estimation technique may be
instrumental in attempts to infer causality in high dimensional phase spaces.

\section{Contributions}
In this paper we have introduced a new multivariate phase distribution,
developed a fast and accurate technique for estimating the parameters of the
distribution, and shown that our technique uniquely recovers the parameters of
the distribution from observations. We have also examined the performance of
the algorithm as a function of the dimensionality and the number of samples,
and provided a demonstration of how this technique could be used for the
recovery of neural coupling in cortex.

\subsubsection*{Acknowledgments}
We would like to thank Bruno Olshausen and Jascha Sohl-Dickstein for helpful
discussions and Mark Timme for comments on the manuscript. We also thank Ryan
Canolty and Robert T. Knight for permitting the use of their experimental data
used in Fig.~\ref{fig:ecog} and Fernando P{\'e}rez for the script used to plot
the network layouts. This work has been supported by NSF grants IIS-0705939
and IIS-0713657, NGA grant HM1582-05-1-2017. The simulations were computed
using IPython~\cite{Perez2007} and NumPy/SciPy~\cite{Oliphant2007}, the
figures were produced using Matplotlib~\cite{Barrett2005} and
NetworkX~\cite{Hagberg2008}.

\renewcommand\refname{\normalsize References}
\bibliographystyle{unsrt} 
\small 

\end{document}